\documentclass[12pt]{iopart}

\usepackage{graphicx}
\begin{document}

\title[A study on the anomaly of $p$ over $\pi$ ratios in $Au+Au$
collisions with jet quenching] {A study on the anomaly of $p$ over
$\pi$ ratios in $Au+Au$ collisions with jet quenching}

\author{Xiaofang Chen$^{1,2}$, Hanzhong Zhang$^{1,2,3}$, Ben-Wei Zhang$^{1,2}$ and Enke Wang$^{1,2}$}

\address{$^1$Institute of Particle Physics, Huazhong Normal
University, Wuhan 430079, China\\
$^2$Key Laboratory of Quark $\&$ Lepton Physics (Huzhong Normal
University), Ministry of Education, China\\
$^3$Department of Physics, Shandong University, Jinan 250100, China}

\begin{abstract}
The ratios of $p/\pi$ at large transverse momentum in central
$Au+Au$ collisions at RHIC are studied in the framework of jet
quenching based on a next-to-leading order pQCD parton model. It is
shown that theoretical calculations with a gluon energy loss larger
than the quark energy loss will naturally lead to a smaller $p/\pi$
ratios at large transverse momentum in $Au+Au$ collisions than those
in $p+p$ collisions at the same energy. Scenarios with equal energy
losses for gluons and quarks and a strong jet conversion are both
explored and it is demonstrated in both scenarios $p/\pi$ ratios at
high $p_T$ in central $Au+Au$ collisions are enhanced and the
calculated ratios of protons over pions approach to the experimental
measurements. However, ${\bar p}/p$ in the latter scenario is found
to fit data better than that in the former scenario.
\end{abstract}

\pacs{12.38.Mh,24.85.+p,25.75.-q}

\maketitle

\section{Introduction}

Relativistic heavy-ion collisions provide the possibility of
creating in the laboratory a plasma of deconfined quarks and gluons,
or QGP. So far a large amount of interesting data have been
accumulated in the experiments at the Relativistic Heavy Ion
Collider (RHIC) \cite{rhic1,rhic2,rhic3,rhic4}, which strongly
suggest a new kind of matter may have been formed at RHIC. One of
the most striking evidences of the formation of QGP is the strong
suppression of hadron spectra with large transverse momentum (or
high $p_T$) in $Au+Au$ collisions as compared to $p+p$
collisions~\cite{star-suppression,phenix-suppression}. This kind of
suppression agrees with the prediction of jet quenching
theory~\cite{quenching-1992,theory1,theory2,theory3,theory4}, which
has provided compelling explanations on many novel phenomena at high
$p_T$ observed at RHIC, such as the suppression of $\pi^0$
production at high $p_T$, and
 the disappearance of back-to-back correlations of high $p_T$
hadrons~\cite{back-to-back}.

In the picture of jet quenching or parton energy loss, an energetic
parton will loss energy by multiple scattering in the medium when it
propagates through the hot QGP medium. The majority of current
approaches to the medium-induced parton energy loss may be divided
into four major schemes which included higher
twist(HT)~\cite{HT1,HT2,HT3,HT4}, path integral approach to the
opacity expansion
(BDMPS-Z/ASW)~\cite{BDMPS1,wiedemann,BDMPS3,BDMPS4}, finite
temperature field theory approach(AMY)~\cite{AMY1,AMY2,AMY3} and
reaction operator approach to the opacity expansion
(GLV)~\cite{GLV1,GLV2,GLV3,GLV4}. It has been shown that the energy
loss of light quarks and gluons is proportional to the gluon density
and has a quadratic dependence on the total distance traversed by
the propagating parton due to the non-Abelian LPM interference
effect in multiple
scatterings~\cite{wiedemann,GLV1,why-quenching,energyloss1,energyloss2}.
Moreover it is expected that the gluon energy loss in nuclei is
$9/4$ times of the quark energy loss due to the different color
factors for quark-gluon vertex and gluon-gluon
vertex~\cite{quenching-1992,energyloss1,energyloss2}. Because high
$p_T$ proton (anti-proton) production in $p+p$ collisions is
dominated by gluon fragmentation whereas high $p_T$ pion at $p+p$
collisions is dominated by quark fragmentation, it is expected that
larger gluon energy loss than quark energy loss in hot/dense medium
will lead to a smaller $p/\pi$ ratio at high $p_T$ in $Au+Au$
collisions than in $p+p$ collisions at the same energy
\cite{ratio-xinnian}.

However, STAR Collaboration recently published very interesting
measurements which indicate that $p/\pi^+$ and $\bar{p}/\pi^-$
ratios at high $p_T$ in central Au+Au collisions~\cite{star-auau}
approach those in $p+p$ and $d+Au$ collisions~\cite{star-pp}, and
the nuclear modification factor of protons is similar to that of
pions \cite{star-auau}. To solve the discrepancy between the
experimental data and the theoretical calculations, a jet conversion
mechanism ~\cite{conversion} has been proposed by taking into
account the possibility of a quark jet in QGP converted into a gluon
jet and vice versa, which has previously been considered
theoretically in Refs. \cite{q-q scattering} for eA deeply inelastic
scattering within higher twist expansion approach. It is found that
conversions between quark and gluon jets indeed lead to an increase
in the final number of gluon jets in central heavy ion collisions
than the case without conversions, but to explain experimental data
observed by STAR a conversion enhancement factor $K=4\sim6$ would be
needed ~\cite{conversion}.

In this paper, we will study the $p/\pi$ ratio in $Au+Au$ collisions
with $\sqrt{s}=200$ GeV by using a next-to-leading order (NLO) pQCD
inspired parton model, which has been very successful in describing
the high $p_T$ region physics in $p+p$ collisions~\cite{Owens}.
Later by including the nuclear effects it is extended to consider
the single hadron and di-hadron productions in $Au+Au$ collisions
and has given a very good description on the high $p_T$ pion and
charged hadron productions at RHIC ~\cite{zhanghz}. An interesting
observation in the model shows that the ratio of gluon/quark jets in
NLO is larger than in LO calculations ~\cite{zhanghz}. In the same
framework, we will investigate the high $p_T$ pion and proton
(anit-proton) spectra, especially the $p/\pi$ ratio at
$\sqrt{s}=200$GeV. We find that with gluon energy loss larger than
quark energy loss in QGP, the NLO pQCD parton model shows the ratios
$p/\pi$ in $Au+Au$ will be smaller than those in $p+p$ collisions,
which agrees with the previous pioneer study in leading-order parton
model~\cite{ratio-xinnian}. To explain the observed $p/\pi$ ratios
at STAR Collaboration phenomenologically, two different scenarios in
jet quenching theory are considered: one is the case that gluon
energy loss is equal to quark energy loss in nuclear medium; another
with a strong jet conversion mechanism where a appreciable quark
jets are converted into gluon jets. Both scenarios will effectively
increase the total number of final gluon jets passing through the
hot QGP medium, and thus give a higher $p/\pi$ ratios, which are
found to reproduce the experimental data fairly well. However,
${\bar p}/p$ in the latter scenario is found to fit data better than
that in the former scenario. In addition, $R_{AA}^p=R_{AA}^{\pi}$ is
found in the former scenario while $R_{AA}^p<R_{AA}^{\pi}$ in the
latter scenario.

The paper is organized as follows. In Sec.~\ref{sec:a}, we describe
explicitly the NLO pQCD parton model in $p+p$ and $Au+Au$
collisions. The effect due to equal energy losses of the quark and
gluon is taken into account in Sec.~\ref{sec:b}. In Sec
.~\ref{sec:c}, the mechanism of flavor conversion between quark and
gluon jet is considered. A reasonable conversion rate with which the
results are in good agreement with experimental data is fixed on.
Finally, we conclude in Sec.~\ref{sec:d} with a summary of present
work.

\section{High $p_T$ hadron production given by a NLO pQCD parton model}\label{sec:a}

PQCD parton model has been shown to work well for large $p_T$
particle production in high energy nucleon-nucleon
collisions\cite{owens1987}. The factorization theorem demonstrates
that the inclusive particle production cross section in $p+p$
collisions can be expressed as a convolution of parton distribution
functions inside the hadron, elementary parton-parton scattering
cross sections and parton fragmentation functions,
\begin{eqnarray}
\frac{d\sigma^{h}_{pp}}{d\Gamma} &=& \sum_{abcd(e)}\int dx_a dx_b
dz_{c(d,e)}f_{a/p}(x_a, \mu^2) f_{b/p}(x_b, \mu^2)\nonumber \\
&~& \times \frac{1}{2 x_a x_b s}\Psi(ab\rightarrow cd(e))
 D_{h/{c(d,e)}}(z_{c(d,e)},\mu^2),\label{eqn:pp}
\end{eqnarray}
where $\Psi(ab\rightarrow cd(e))$ is related to the squared matrix
elements for the various $2 \rightarrow 2$ or $2 \rightarrow 3$
subprocesses in parton-parton hard scattering. $d \Gamma$ is the
differential two-body or three-body phase space element. The
detailed expressions and introductions for $\Psi(ab\rightarrow
cd(e))$ and $d \Gamma$ can be found in
references~\cite{Owens,Soper}. $f_{a/p}(x_a,\mu^2)$ is the normal
parton distribution function (PDF) for which we will use the CTEQ6M
parametrization \cite{distribution}, and $x_a$ is the fraction of
the hadron's momentum carried by the parton. $D^0_{h/c}(z_c,\mu^2)$
is the fragmentation function (FF) of parton $c$ into hadron $h$,
and $z_c$ is the momentum fraction of a parton jet carried by a
produced hadron. We will use mainly the updated AKK
parametrization~\cite{akk08} for jet fragmentation, which has
recently been improved from both new theoretical input and RHIC
data, especially for $\pi^\pm$, $K^\pm$, $p/\bar{p}$, $K_S^0$ and
$\Lambda/\bar{\Lambda}$ particles.

\begin{figure}
\begin{center}
\includegraphics[width=75mm]{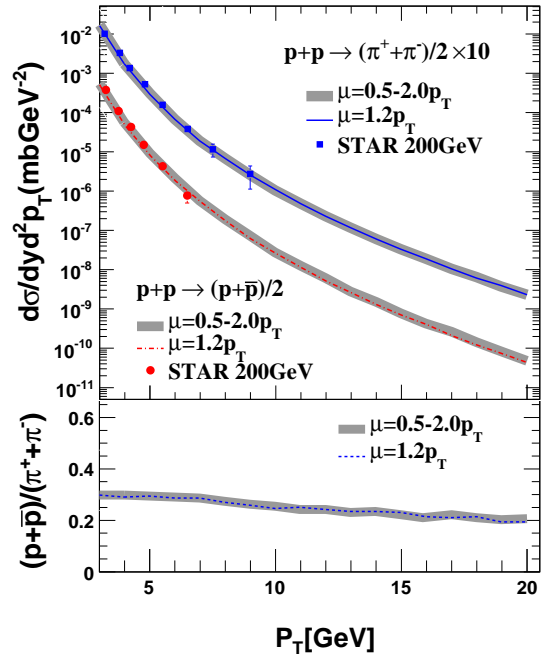}
\end{center}
    \caption{(Color online) The hadron spectra for
    $(\pi^{+}+\pi^{-})/2$ and $(p+\bar{p})/2$ and the ratios of $(p+\bar{p})/(\pi^{+}+\pi^{-})$
    in $p+p$ collisions at $\sqrt{s}$ = 200 GeV. Different values of the scales for factorization, renormalization and fragmentation
    are used in numerical calculations. The data are from Ref\cite{star-auau}.}
     \label{fig:pp}
\end{figure}

The calculations discussed in this paper are carried out within a
NLO Monte Carlo based program which uses a variant of the phase
space slicing technique~\cite{Owens,bergmann}. This allow the same
kinematic cuts in the extraction of the data to be imposed on the
theoretical predictions. For the lowest-order contributions of
$2\rightarrow2$ tree level ($\alpha_s^2$ order), the NLO correction
contributions ($\alpha_s^3$ order) include 1-loop correction
contributions to $2\rightarrow2$ tree level and $2\rightarrow3$ tree
level contributions, which utilize two-cutoff
parameters~\cite{Owens}, $\delta_s$ and $\delta_c$, to isolate the
soft and collinear divergences in the squared matrix elements. The
results in a set of two-body and three-body weights depend on the
cut-offs, but this dependence cancels when the weights are combined
in the calculation of physical observations.

For the calculations of the cross section at fixed values of the
transverse momentum $p_T$ of the produced hadron, we will choose the
factorization scale, the renormalization scale and the fragmentation
scale $\mu$ to be all proportional to $p_T$. As pointed out in
Ref.\cite{zhanghz}, the calculated single inclusive $\pi^0$ spectra
in the NLO pQCD parton model in $p+p$ collisions agree well with the
experimental data at the RHIC energy with the scale in the range
$\mu=0.9 \sim 1.5 p_T$. Shown in Fig.~(\ref{fig:pp}) we give the
spectra for charged pion and (anti-)proton in $p+p$ collisions and
their ratios $(p+\bar{p})/(\pi^{+}+\pi^{-})$ as functions of $p_T$.
The grey bands are with scales $\mu=0.5-2.0p_T$. Not only the NLO
spectra but also the ratios in $p+p$ collisions are not very
sensitive to the scale. So we will choose the same scale
$\mu=1.2p_T$ in the following calculations for both $p+p$ and
$Au+Au$ collisions.

\begin{figure*}[t]
\begin{center}
\includegraphics[width=4.5in]{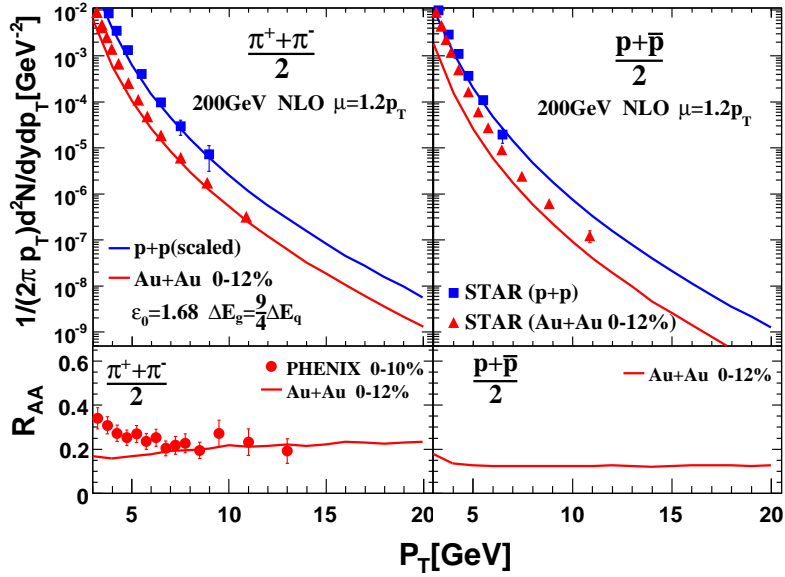}
\end{center}
\caption{(Color online) The spectra of $(\pi^{+}+\pi^{-})/2$ and
$(p+\bar{p})/2$ in $p+p$ and $Au+Au$ collisions, and the suppression
factors in central $Au+Au$ collisions at $\sqrt{s} = 200 GeV$. The
data are from Ref.\cite{star-auau}.} \label{fig:ppaa-spec1}
\end{figure*}

Assuming the $N+N$ cross section can be extrapolated to $A+A$
collisions, the inclusive invariant differential cross section for
producing a hadron $h$ in a collision of nucleuses of types $A$ and
$B$ can be written as,
\begin{eqnarray}
\frac{d\sigma^{h}_{AA}}{d\Gamma} &=&
\sum_{abcd(e)}\frac{1}{2\pi}\int d\phi d{b^2}d{r^2}dx_a dx_b
dz_{c(d,e)}\nonumber \\
&~& \times t_A({\bf r})t_A(|{\bf r}-{\bf b}|)\nonumber \\
&~& \times f_{a/A}(x_a, \mu^2,{\bf
r}) f_{b/A}(x_b, \mu^2,|{\bf r}-{\bf b}|)\nonumber \\
&~& \times \frac{1}{2 x_a x_b s}\Psi(ab\rightarrow cd(e))\nonumber \\
&~& \times  D_{h/{c(d,e)}}(z_{c(d,e)},\mu^2,\Delta
E_{c(d,e)})\,,\label{eq:AA}
\end{eqnarray}
where $\phi$ is the azimuthal angle with respect to the reaction
plane and describes the direction a final parton jet is emitted in
the overlap region between two colliding nuclei. The function
$t_A(r)=\frac{3A}{2\pi R^2}\sqrt{1-r^2/R^2}$ is the nuclear
thickness function in a hard-sphere geometry model, normalized to
$\int d^2rt_A(r)=A$ which is the nucleon number or mass number in a
nuclei with radius $R=1.12A^{1/3}$ fm. $f_{a/A}(x_a,\mu^2,r)$ is
assumed to be factorizable into the parton distribution in a free
nucleon $f_{b/N}(x,\mu^2)$ and the nuclear shadowing factor
$S_{a/A}(x,{\bf r})$ given by HIJING
parameterization\cite{shadowing},
\begin{eqnarray}
   f_{a/A}(x,\mu^2,\mathbf{r})&=&S_{a/A}(x,\mathbf{r})
   [\frac{Z}{A}f_{a/p}(x,\mu^2)\nonumber \\
   &~&+(1-\frac{Z}{A})f_{a/n}(x,\mu^2)]\,,
   \label{eq:distribution}
\end{eqnarray}
where we have explicitly taken into account the isospin of the
nucleus by considering the parton distributions of a neutron which
are obtained from that of a proton by isospin symmetry.

As assumed in Ref. \cite{why-quenching,zhanghz}, the effect of
final-state interaction between produced parton and the bulk medium
can be described by the effective medium-modified FF's,
\begin{eqnarray}
D_{h/c}(z_c,\Delta E_c,\mu^2) &=&(1-e^{-\langle \frac{L}
{\lambda}\rangle}) \left[ \frac{z_c^\prime}{z_c}
D^0_{h/c}(z_c^\prime,\mu^2) \right.
 \nonumber \\
&~& \left. + \langle \frac{L}{\lambda}\rangle \frac{z_g^\prime}{z_c}
D^0_{h/g}(z_g^\prime,\mu^2)\right] \nonumber \\
&~&+ e^{-\langle\frac{L}{\lambda}\rangle} D^0_{h/c}(z_c,\mu^2),
\label{eq:modfrag}
\end{eqnarray}
where $z_c^\prime=p_T/(p_{Tc}-\Delta E_c)$, $z_g^\prime=\langle
L/\lambda\rangle p_T/\Delta E_c$ are the rescaled momentum
fractions, $\Delta E_c$ is the average radiative parton energy loss
and $\langle L/\lambda\rangle$ is the number of scatterings along
the parton propagating path,
\begin{equation}
\langle L/\lambda\rangle =\int_{\tau_0}^{\infty}
\frac{d\tau}{\rho_0\lambda_0} \rho_g(\tau,{\bf b},{\bf r}+{\bf
n}\tau).
\end{equation}

Neglecting transverse expansion, the gluon density distribution in a
1-d expanding medium in $A+A$ collisions at impact-parameter ${\bf
b}$ is assumed to be proportional to the transverse profile of
participant nucleons,
\begin{equation}
\rho_g(\tau,{\bf b},{\bf r})=\frac{\tau_0\rho_0}{\tau}
        \frac{\pi R^2_A}{2A}[t_A({\bf r})+t_A(|{\bf b}-{\bf r}|)].
\end{equation}

\begin{figure*}[t]
\begin{center}
\includegraphics[width=4.in]{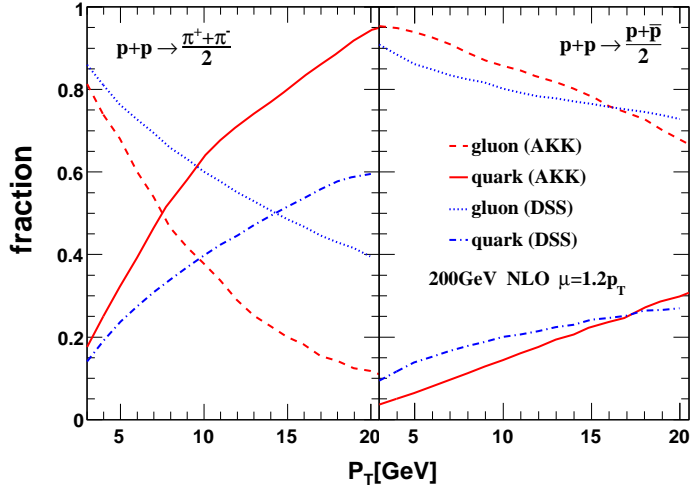}
\end{center}
    \caption{(Color online) The fractions of contributions to the charged $\pi$
    and (anti-)proton from quark and gluon jets with AKK fragmentation functions and
    DSS fragmentation functions in $p+p$ collisions at $\sqrt{s}=200$ GeV, respectively.}
     \label{fig:fraction}
\end{figure*}

According to recent theoretical studies
\cite{theory4,wiedemann,2-wang} the total parton energy loss in a
finite and expanding medium can be approximated as a path integral,
\begin{equation}
\Delta E \approx \langle \frac{dE}{dL}\rangle_{1d}
\int_{\tau_0}^{\infty} d\tau \frac{\tau-\tau_0}{\tau_0\rho_0}
\rho_g(\tau,{\bf b},{\bf r}+{\bf n}\tau),
\end{equation}
for a parton produced at a transverse position ${\bf r}$ and
traveling along the direction ${\bf n}$. $\langle dE/dL\rangle_{1d}$
is the average parton energy loss per unit length in a 1-d expanding
medium with an initial uniform gluon density $\rho_0$ at a formation
time $\tau_0$ for the medium gluons. The energy dependence of the
energy loss is parameterized as
\begin{equation}
 \langle\frac{dE}{dL}\rangle_{1d}=\epsilon_0 (E/\mu_0-1.6)^{1.2}
 /(7.5+E/\mu_0),
\label{eq:loss}
\end{equation}
from the numerical results in Ref.~\cite{2-wang} in which thermal
gluon absorption is also taken into account. The parameter
$\epsilon_0$ should be proportional to the initial gluon density
$\rho_0$. A simultaneous fit to the single and dihadron data
constrains the energy loss parameter within a narrow range:
$\epsilon_0=1.6-2.1$ GeV/fm in Ref.\cite{zhanghz}. For the following
calculations we will choose the same parameters as in the reference,
$\mu_0=1.5$ GeV, $\epsilon_0\lambda_0=0.5$ GeV and $\tau_0=0.2$ fm
in $Au+Au$ collisions at $\sqrt{s}$ = 200 GeV.

Shown in Fig.~(\ref{fig:ppaa-spec1}) are the charged $\pi$ and
(anti-)proton spectra and the suppression factors in central $Au+Au$
collisions at $\sqrt{s}$ = 200 GeV. The suppression factor (or
nuclear modification factor) which mainly manifests jet quenching
effect in our current studies is defined as~\cite{modification
factor},
\begin{eqnarray}
   R_{AA}=\frac{d\sigma_{AA}^h/dyd^2p_T}{\langle N_{binary}\rangle
   d\sigma_{pp}^h/dyd^2p_T},
   \label{eq:rab}
\end{eqnarray}where $N_{binary}$ is the average number of geometrical binary
collisions at a given range of impact parameters, $\langle
N_{binary}\rangle=\int d^2bd^2r t_A(r)t_A(|\mathbf{b}-\mathbf{r}|)$.

Non-Abelian feature of QCD show that gluon energy loss is $9/4$
times of quark energy loss in the dense matter due to the difference
of color factors for gluon-gluon vertex and quark-gluon
vertex~\cite{quenching-1992,energyloss1,energyloss2}. With the
parton energy loss, $\Delta E_g=9/4\Delta E_q$, and the energy loss
parameter chosen as $\epsilon_0$ = 1.68 GeV/fm, the calculated large
transverse momentum $(\pi^++\pi^-)/2$ spectra in central $Au+Au$
collisions fit data well and the large transverse momentum $(p+\bar
p)/2$ spectra fit data bad in Fig.~(\ref{fig:ppaa-spec1}). Also
shown in the figure are the suppression factors for
$(\pi^++\pi^-)/2$ and $(p+\bar p)/2$ spectra in central $Au+Au$
collisions, respectively. The former is found larger than the
latter.

Shown in Fig.(\ref{fig:fraction}) are the fractions of the
contributions to the charged pion and (anti-)proton from quark and
gluon jets with different fragmentation functions, respectively. In
the two plots the red solid curves and the red dashing curves are
with AKK fragmentation functions\cite{akk08} while the blue dotted
curves and the blue dashed-dotted curves are with DSS fragmentation
functions\cite{DSS}. With the two sets of fragmentation functions,
our NLO numerical results show that high $p_T$ (anti-)proton is
dominated by gluon fragmentation whereas high $p_T$ charged pion is
dominated by quark fragmentation. It is consistent with a previous
study based on Monte Carlo simulations\cite{ratio-xinnian}.

\begin{figure}
\begin{center}
\includegraphics[width=80mm]{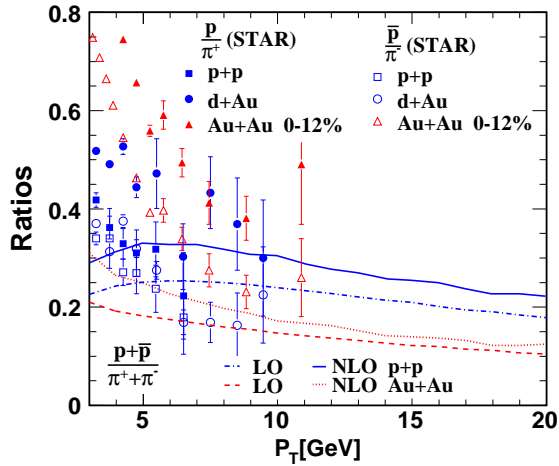}
\end{center}
    \caption{(Color online) The ratios of $(p+\bar{p})/(\pi^{+}+\pi^{-})$
    in $p+p$ and 0-12\% $Au+Au$ collisions at $\sqrt{s}$ = 200
   GeV. The energy loss parameter is chosen as $\epsilon_0$ = 1.68
   GeV/fm. The gluon energy loss is considered as about 2 times of the quark energy loss.}
     \label{fig:ratio-p-pie}
\end{figure}

Because the gluon energy loss is considered as $9/4$ times of the
quark energy loss, the contribution from gluon jets is suppressed
more greatly than the contribution from quark jets in central
$Au+Au$ collisions. The non-Abelian feature of parton energy loss
causes to a smaller suppression factor $R_{AA}$ for large transverse
momentum $(p+\bar p)/2$ than that for $(\pi^++\pi^-)/2$, as shown in
Fig. (\ref{fig:ppaa-spec1}). So a smaller $p/\pi$ ratio at high
$p_T$ in central $Au+Au$ collisions than that in $p+p$ collisions is
obtained, as shown in Fig. (\ref{fig:ratio-p-pie}). These
conclusions had been previously pointed out in Ref.
\cite{ratio-xinnian} within LO pQCD parton model.

Since the number ratio of gluon/quark jets created in hard
scattering processes in NLO calculations is larger than in LO
calculations \cite{zhanghz}, the contribution fraction from gluon
jets will be enhanced in NLO compared with that in LO. Therefore,
the NLO $p/\pi$ ratio is larger than the LO ratio in both $p+p$ and
central $Au+Au$ collisions. However, even in NLO calculations the
$p/\pi$ ratio in central $Au+Au$ collisions is still smaller than
that in $p+p$ collisions, as shown in Fig.(\ref{fig:ratio-p-pie}),
while recent STAR data indicate that $p/\pi^+$ and $\bar{p}/\pi^-$
ratios at high $p_T$ in central $Au+Au$ collisions~\cite{star-auau}
approach those in $p+p$ and $d+Au$ collisions~\cite{star-pp}. To
solve this discrepancy between the theoretical calculations and
experiment measurements, we will explore two scenarios: one is that
gluon jets lose the same amount of energy as quark jets; another is
jet conversion between gluon and quark jets. In the following we
demonstrate that how  $p/\pi$  and $p/\bar{p}$ ratios will be
modified in these two scenarios, and their comparisons with the
experimental data.

\section{Scenario I: gluon jet energy loss is equal to quark jet energy loss}\label{sec:b}

\begin{figure}
\begin{center}
\includegraphics[width=80mm]{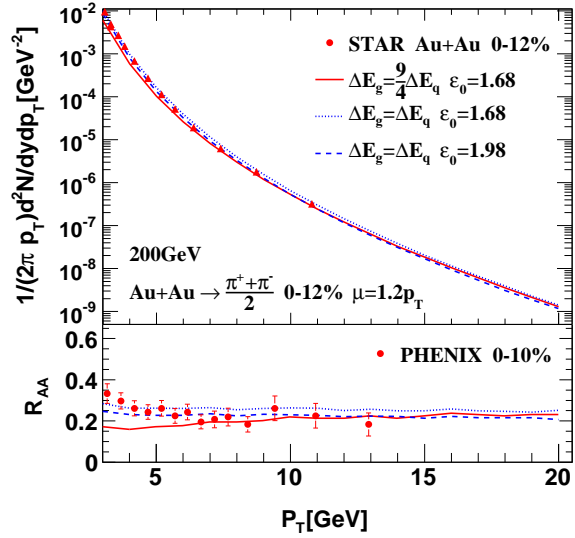}
\end{center}
    \caption{(Color online) The charged $\pi$ meson spectra (upper) and the suppression factors
    (lower) with 2 sets of parton energy loss
    ($\Delta E_g=9/4\Delta E_q$ and $\Delta E_g=\Delta E_q$) in central
    $Au+Au$ collisions at $\sqrt{s}$ = 200 GeV.
    The data points are from \cite{star-pp,phenix}.}
     \label{fig:spec-pi-equalenergyloss}
\end{figure}

\begin{figure*}
\begin{center}
\includegraphics[width=4.5in]{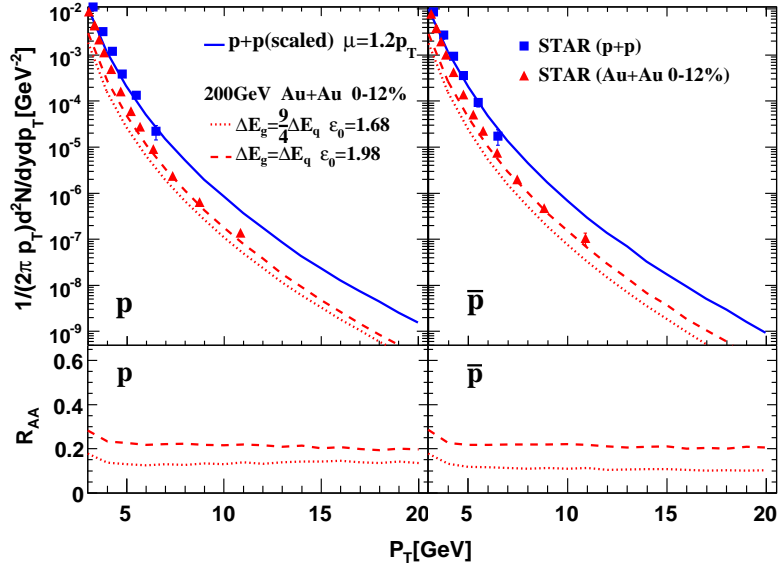}
\end{center}
    \caption{(Color online) The separated $p$ and $\bar{p}$ spectra and the suppression factors in
    central $Au+Au$ collisions at $\sqrt{s}$ = 200 GeV.
    The data are from \cite{star-pp,phenix}.}
     \label{fig:spec-proton-equalenergyloss}
\end{figure*}

To explain phenomenologically the observed $p/\pi$ ratios given by
STAR Collaboration, we first consider a case in jet quenching
theory: equal energy loss for both gluon and quark jets. Though
currently there is no such a theory of jet quenching to justify that
gluons will lose the same amount of energy as quarks, this toy model
could shed light on the flavor dependence of parton energy loss in
nuclear. We note such a scenario of parton energy loss had been
explored in a study of the sensitivity of hadron spectra suppression
to the non-Abelian parton energy loss~\cite{non-abel}.

The hadron yield will be less suppressed with parton energy loss
$\Delta E_g=\Delta E_q$ than that with parton energy loss $\Delta
E_g=\frac{9}{4}\Delta E_q$ in central $Au+Au$ collisions, so one
should enlarge the energy loss parameter $\epsilon_0$ to fit data.
Shown in Fig.~(\ref{fig:spec-pi-equalenergyloss}) are the numerical
results with 2 sets of parton energy loss ($\Delta
E_g=\frac{9}{4}\Delta E_q$ and $\Delta E_g=\Delta E_q$) for charged
pion meson spectra and the corresponding suppression factors
compared with data. From the suppression factors in
Fig.~(\ref{fig:spec-pi-equalenergyloss}), it is clear that
$\epsilon_0$ = 1.98 GeV/fm works well for fitting data. The
parameter value will be used to give (anti-)proton spectra with
parton energy loss $\Delta E_g=\Delta E_q$ in central $Au+Au$
collisions.

AKK parametrization of fragmentation functions just give parton
fragmentation functions to $(\pi^+ + \pi^-)/2$ or $(p + \bar{p})/2$.
To get ratios  $p/\pi^+$, $\bar{p}/\pi^-$ and $p/\bar{p}$ we need
the parton fragmentation function to $\pi^+$, $\pi^-$, $p$ and
$\bar{p}$ respectively. Theoretical and experimental
studies~\cite{star-auau,star-pp,conversion,straub} show that the
fragmentation of quark and anti-quark jets produce mainly protons
and anti-protons, respectively, and gluon jets contribute to protons
and anti-protons equally. As in Ref.~\cite{star-pp,conversion}, we
assume that no anti-protons are produced from a quark jet and no
protons are produced from an anti-quark jet. We further assume that
high $p_T$ charged $\pi^+$ and $\pi^-$ mesons are equally produced
from hard parton jet fragmentations, consistent with experiment
findings~\cite{star-auau,star-pp}.

Based on the NLO pQCD parton model, we calculate the separated
proton and anti-proton spectra with the energy loss parameter
$\epsilon_0$ = 1.98 GeV/fm and the equal energy loss for both gluon
and quark jets in central $Au+Au$ collisions at $\sqrt{s}$ = 200
GeV. The numerical results are shown in
Fig.~(\ref{fig:spec-proton-equalenergyloss}) where the suppression
factors are also included. The equal energy loss for both gluon and
quark jets for (anti-)proton production in central $Au+Au$
collisions results in the suppression factor $R_{AA}^p\approx0.2$,
equal to $R_{AA}^{\pi}\approx0.2$ shown in
Fig.~(\ref{fig:spec-pi-equalenergyloss}). As shown in
Fig.~(\ref{fig:fraction}), high $p_T$ (anti-)proton production is
dominated by gluon fragmentation. Compared with the case of parton
energy loss $\Delta E_g=\frac{9}{4}\Delta E_q$, gluon jets with less
energy loss $\Delta E_g=\Delta E_q$ may have larger probability to
pass through the hot medium and this gives a larger percentage of
contribution for (anti-)proton production, so one can see in
Fig.~(\ref{fig:spec-proton-equalenergyloss}) that (anti-)proton
spectra with parton ernergy loss $\Delta E_g=\Delta E_q$ are
enhanced although the energy loss parameter $\epsilon_0$ is adjusted
large to 1.98 GeV/fm instead of 1.68 GeV/fm.

\begin{figure}
\begin{center}
\includegraphics[width=75mm]{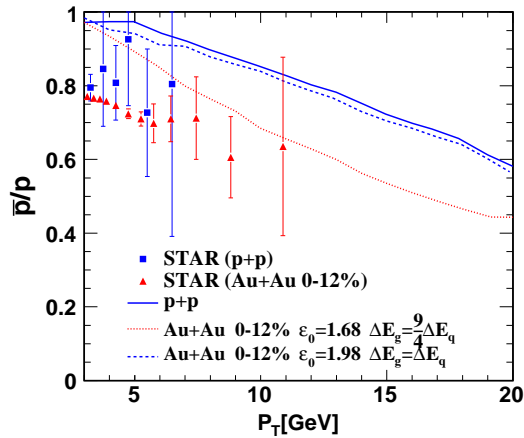}
\end{center}
    \caption{(Color online) The $\bar{p}/p$ ratios at
    mid-rapidity ($|y|<0.5$) as functions of $p_T$ in $p+p$ (solid curve) and $0-12\%$
    $Au+Au$ collisions at $\sqrt{s}$ = 200 GeV. The data are from \cite{star-auau,star-pp}.}
    \label{ratio-pbar-p-equalenergyloss}
\end{figure}

Shown in Fig.~(\ref{ratio-pbar-p-equalenergyloss}) are $\bar{p}/p$
ratios in $p+p$ and central $Au+Au$ collisions at $\sqrt{s}$ = 200
GeV. According to our NLO calculations on the fractions of
contributions to (anti-)proton from the fragmentaions of quark and
gluon jets shown in Fig.~(\ref{fig:fraction}), one know that the
ratio of gluon to quark jet contributing to (anti-)proton production
decreases with the (anti-)proton transverse momentum. Because most
of antiprotons come from gluons while protons come from both valence
quark and gluon fragmentation~\cite{ratio-xinnian}, the ratio of
antiproton to proton production cross section should also decrease
with their $p_T$ , as our NLO calculations show in Fig.
(\ref{ratio-pbar-p-equalenergyloss}). Similar to the LO calculations
in Ref. \cite{ratio-xinnian}, the ratio will always be smaller than
1 because there are much more quark jets than anti-quark jets
produced in hard scattering in $p+p$ collisions due to baryon number
conservation. In central $Au+Au$ collisions the non-Abelian feature
of parton energy loss ($\Delta E_g=9/4\Delta E_q$) reduces the
contribution proportion from gluon jet fragmentations compared with
that in $p+p$ collisions, so the large transverse momentum
$\bar{p}/p$ ratio in central $Au+Au$ collisions is found to be
smaller than that in $p+p$ collisions. If gluon jets lose less
energy ($\Delta E_g=\Delta E_q$ instead of $\Delta E_g=9/4\Delta
E_q$), the contribution from gluon jet fragmentations is enhanced,
therefore there is a larger $\bar{p}/p$ ratio with $\Delta
E_g=\Delta E_q$ than that with $\Delta E_g=9/4\Delta E_q$ in central
$Au+Au$ collisions. In another point of view, equal energy loss for
both quark and gluon jets causes to equal suppression for both $p$
and $\bar{p}$ spectra (shown in
Fig.~(\ref{fig:spec-proton-equalenergyloss})), so the large
transverse momentum $\bar{p}/p$ ratio with $\Delta E_g=\Delta E_q$
in central $Au+Au$ collisions is equal to that in $p+p$ collisions.

\begin{figure*}
\begin{center}
\includegraphics[width=4.5in]{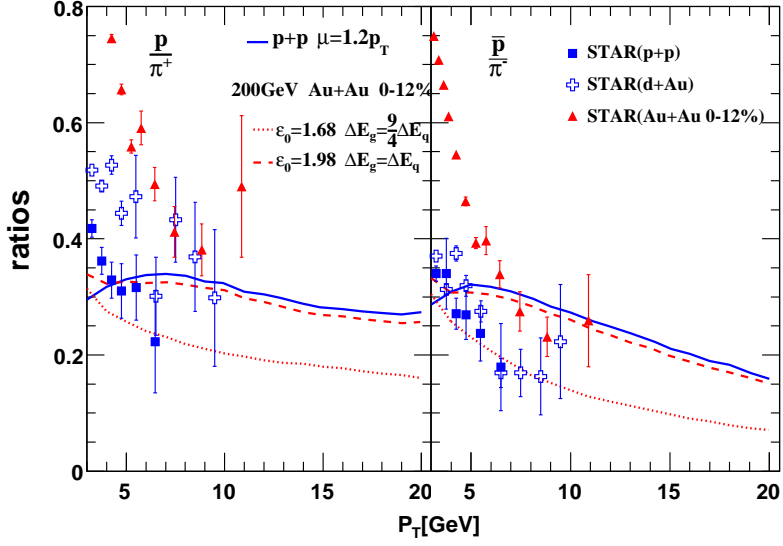}
\end{center}
\caption{(Color online) The $p/\pi^{+}$ and $\bar{p}/\pi^{-}$ ratios
at mid-rapidity($|y|<0.5$) as functions of $p_T$ in $p+p$ collisions
(solid line) and $0-12\%$ $Au+Au$ collisions at $\sqrt{s}$ = 200
GeV. The data are from \cite{star-auau,star-pp}.}
\label{fig:ratio-p-pie-equalenergyloss}
\end{figure*}

In Fig.~(\ref{fig:ratio-p-pie-equalenergyloss}) we make a comparison
of $p(\bar{p})/\pi^{+}(\pi^{-})$ ratios  with 2 sets of parton
energy loss ($\Delta E_g=9/4\Delta E_q$ and $\Delta E_g=\Delta E_q$)
in central $Au+Au$ collsions at $\sqrt{s}$ = 200 GeV. On one hand,
if gluon jets lose less energy ($\Delta E_g=\Delta E_q$ instead of
$\Delta E_g=9/4\Delta E_q$), more gluon jets are survived to
contribute to hadron production when they encounter multiple
scattering in the hot and dense matter created in central $Au+Au$
collisions. The gluon contribution has an enhanced fraction while
the quark contribution has a reduced fraction, and because high
$p_T$ proton (anti-proton) production is dominated by gluon
fragmentation whereas high $p_T$ pion is dominated by quark
fragmentation, the $p(\bar{p})/\pi^{+}(\pi^{-})$ ratios with $\Delta
E_g=\Delta E_q$ are larger than those with $\Delta E_g=9/4\Delta
E_q$ shown in Fig.~(\ref{fig:ratio-p-pie-equalenergyloss}). On the
other hand, equal energy loss for both quark and gluon jets washes
out the suppression difference between (anti-)proton and pion
spectra, $R_{AA}^p(p_T) \approx R_{AA}^{\pi}(p_T)$, shown in
Fig.~(\ref{fig:spec-pi-equalenergyloss}) and
(\ref{fig:spec-proton-equalenergyloss}). In fact, the $p/\pi$ ratio
with parton energy loss $\Delta E_g=\Delta E_q$ in central $Au+Au$
collisions can be written as
\begin{eqnarray}
(\frac{p}{\pi})^{AuAu}=(\frac{R_{AA}^p}{R_{AA}^{\pi}})(\frac{p}{\pi})^{pp}
\approx(\frac{p}{\pi})^{pp}. \label{eq:modfrag}
\end{eqnarray}
That's why our numerical results for large transverse momentum
$p/\pi$ ratios with $\Delta E_g=\Delta E_q$ in central $Au+Au$
collisions approach to those in $p+p$ collisions, consistent with
experiment findings\cite{star-auau,star-pp} shown in
Fig.~(\ref{fig:ratio-p-pie-equalenergyloss}).

\section{Scenario II: Strong Conversion between hard quark and gluon jets}\label{sec:c}

\begin{figure}
\begin{center}
\includegraphics[width=80mm]{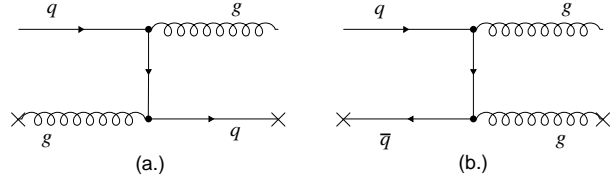}
\end{center}
\caption{ Conversions between quark and gluon jets via both elastic
$gq(\bar q)\to q(\bar q)g$ and inelastic $q\bar q\leftrightarrow gg$
scatterings with thermal quarks and gluons in the hot and dense
matter.}
 \label{feynman-conversion}
\end{figure}

The conversion between quark and gluon jets is a mechanism
introduced in Ref.\cite{conversion,q-q scattering} to explain the
discrepancy for $p/\pi$ ratios between the experimental data and the
theoretical calculations. Although a conversion enhancement factor
$K=4\sim6$ would be needed to explain experimental data observed by
STAR and further theoretical studies will be needed to understand
how we can get  such a large conversion enhancement factor, this
mechanism indeed leads to an increase in the final number of gluon
jets in central $Au+Au$ collisions than the case without conversions
~\cite{conversion}, and thus provide a promising way to explain the
striking experimental data on proton over pion ratio observed at
STAR collaboration~\cite{star-auau}.
In this section we will apply the jet conversion mechanism into the
NLO pQCD parton model to study (anti-)proton and charged pion
spectra and their ratios in heavy ion collisions.

\begin{figure}
\begin{center}
\includegraphics[width=70mm]{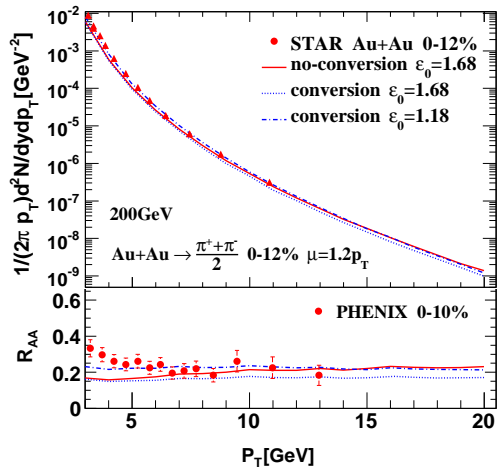}
\end{center}
   \caption{(Color Online) The charged pion spectra and
   nuclear modification factors with and without conversion
   in central $Au+Au$ collisions at $\sqrt{s}=200$ GeV.
    The data are from \cite{star-pp,phenix}.}
\label{spec-pie-conversion}
\end{figure}

\begin{figure*}
\begin{center}
\includegraphics[width=4.5in]{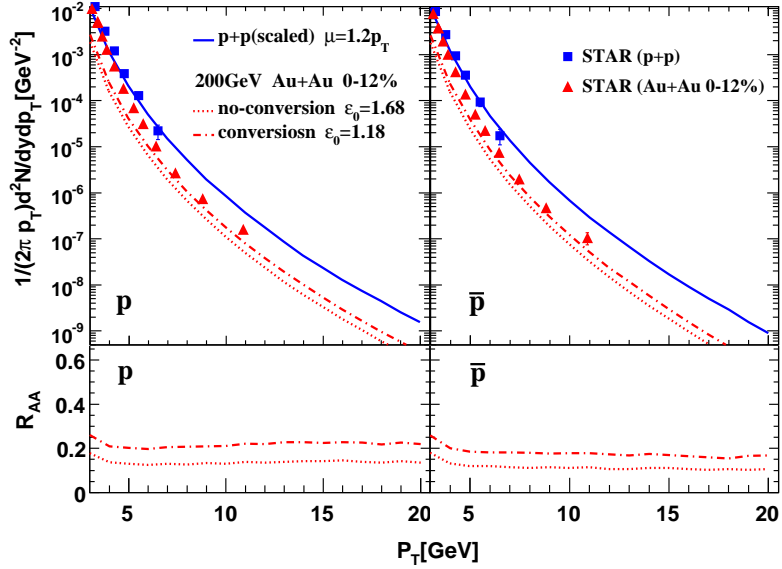}
\end{center}
    \caption{(Color online) The separated $p$ and $\bar{p}$ spectra and the suppression factors
    with and without conversion in
    central $Au+Au$ collisions at $\sqrt{s}$ = 200 GeV.
    The data are from \cite{star-pp,phenix}.}
     \label{fig:spec-proton-conversion}
\end{figure*}

Conversions between quark and gluon jets could happen via both
elastic $gq(\bar q)\to q(\bar q)g$ and inelastic $q\bar
q\leftrightarrow gg$ scatterings with thermal quarks and gluons in
the hot and dense matter as illustrated in
Fig.~(\ref{feynman-conversion}). For a parton jet created at a
spatial point in $Au+Au$ collisions region and escaping out off the
dense matter in the transverse plane, multiple scattering with the
medium particles wears off the jet energy and also stimulates the
jet to convert to its partner. The conversion possibility for the
jet passing through the matter is in turn connected with its energy
and the medium particle density distributed along its passing path.

\begin{figure}
\begin{center}
\includegraphics[width=70mm]{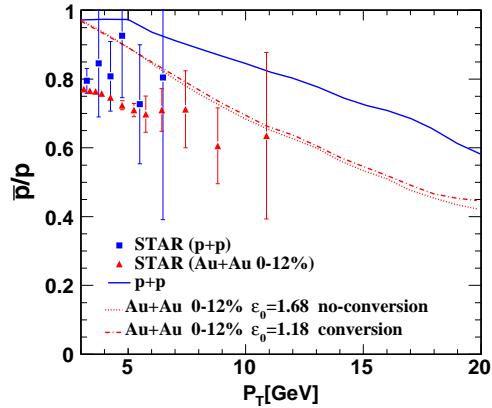}
\end{center}
    \caption{(Color online) The $\bar{p}/p$ ratio at
    mid-rapidity($|y|<0.5$) as functions of $p_T$ in $p+p$ (solid curve)and $0-12\%$
    $Au+Au$ (with and without conversion) collisions at $\sqrt{s}$ = 200 GeV.
    The data are from \cite{star-auau,star-pp}.}
    \label{fig:ratio-pbar-p-conversion}
\end{figure}

As studied in Ref.~\cite{conversion,liu-fries}, the quark-to-gluon
conversion possibility is found to be larger than the gluon-to-quark
possibility at the same jet energy, and about 30\% quarks for net
quark-to-gluon jet conversions are needed to account for the
observed ratios in the experiment in central $Au+Au$ collisions. For
simple evaluation for net quark-to-gluon conversions, we neglect
gluon-to-quark jet conversions, and assume the quark-to-gluon jet
conversion has an averaged possibility 45\% all over its path. In
our actual calculation, the quark and gluon jet contributions to the
observed hadrons can be separated within a NLO pQCD parton model in
A+A collisions. Once quark jets are created anywhere inside the
medium, 45\% quark jets are marked as gluon jets who will encounter
multiple scattering and then fragment into hadrons. After a
quark-to-gluon jet conversion happens at some point on the path in
the medium, the newcomer (gluon) will have energy loss $\Delta
E_g=9/4\Delta E_q$ different from the energy loss $\Delta E_q$ for
its parent jet (quark). Because the ``new" gluon is easier absorbed
by the hot and dense matter than the parent quark, the total jet
number will be reduced by net quark-to-gluon conversions, and there
are actually about 30\% of final surviving conversion jets. The
ratio 30\% is consistent to the studies in
Ref.~\cite{conversion,liu-fries}. In our calculations jet energy
loss depends on the local parton density and total propagation
length. A more sophisticated consideration will be needed in the
further studies for the net quark-to-gluon conversion dependent on
the local parton density and total propagation length.

\begin{figure*}
\begin{center}
\includegraphics[width=4.5in]{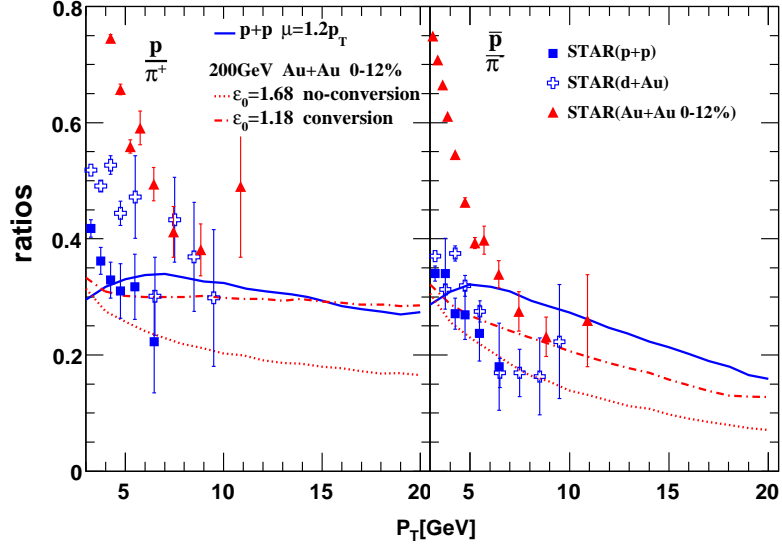}
\end{center}
 \caption{(Color online) The $p/\pi^{+}$ and
$\bar{p}/\pi^{-}$ ratios at mid-rapidity($|y|<0.5$) as functions of
$p_T$ in $p+p$ and central $Au+Au$ collisions with conversion
(dotted-dashed line) and without conversion (dotted line) between
quarks and gluons. Data points are taken from
\cite{star-auau,star-pp}.} \label{fig:ratio-p-pie-conversion}
\end{figure*}

The appearance of net quark-to-gluon jet conversions will increase
the total number of final gluon jets passing through the hot QGP
medium. In such a conversion framework with gluon energy loss larger
than quark energy loss, $\Delta E_g=9/4\Delta E_q$, the high $p_T$
hadron yield should be smaller than that without conversion. To fit
data for hadron spectra, the energy loss parameter $\epsilon_0$
should be adjusted to be smaller than 1.68 GeV/fm.

Shown in Fig.~(\ref{spec-pie-conversion}) are the charged pion
spectra and nuclear modification factors with and without conversion
in central $Au+Au$ collisions at $\sqrt{s}=200$ GeV. Our numerical
results with conversion mechanism and $\epsilon_0$ = 1.18 GeV/fm are
found to fit data well. With this energy loss parameter we then show
in Fig.~(\ref{fig:spec-proton-conversion}) separately $p$ and
$\bar{p}$ spectra and the suppression factors with and without
conversion in central $Au+Au$ collisions. The spectra with
conversion are found to fit data better than without conversion, and
the nuclear modification factor for proton production with
conversion in Fig.~(\ref{fig:spec-proton-conversion}) is similar to
that with conversion for pion production in
Fig.~(\ref{spec-pie-conversion}), $R_{AA}^{p}\approx R_{AA}^{\pi}$
due to the increasing number of converted gluon jets which dominate
proton productions. However, $R_{AA}$ for anti-proton with
conversion is slightly smaller than that for proton with conversion,
$R_{AA}^{\bar p}/R_{AA}^{p}\approx 0.75$, because most of
antiprotons come from gluons with energy loss larger than quark
while protons come from both valence quark and gluon fragmentation.

In Fig.~(\ref{fig:ratio-pbar-p-conversion}) we plot the $\bar{p}/p$
ratio as a function of $p_T$ in $p+p$ (solid curve) and central
$Au+Au$ (with and without conversion) collisions. Of interest is
that the ratio $\bar{p}/p$ with conversion is found to be
approximately equal to that without conversion, and fit data better
than $\bar{p}/p$ in the scenario of equal energy loss for both quark
and gluon jets in central $Au+Au$ collisions. In fact, according to
the result in Fig.~(\ref{fig:spec-proton-conversion}), $R_{AA}^{\bar
p}/R_{AA}^{p}\approx 0.75$, one can obtain $({\bar p}/{
p})_{AuAu}\approx 0.75({\bar p}/{p})_{pp}$, consistent with the
numerical results in Fig.~(\ref{fig:ratio-pbar-p-conversion}). The
net quark-to-gluon jet conversions enhance the total number of the
final gluon jets as compared with no conversion, and most of
antiprotons come from gluons while protons come from both valence
quark and gluon fragmentation, so the ratio $\bar{p}/p$ with
conversion should be smaller than that with no conversion. But
compared with the no conversion case with $\epsilon_0$ = 1.68
GeV/fm, the smaller energy loss parameter $\epsilon_0$ = 1.18 GeV/fm
for the conversion case should give rise to the $\bar{p}/p$ curve
for the conversion case a bit closer to the curve for the $p+p$
case. Therefore, the ratio $\bar{p}/p$ with conversion is found to
be approximately equal to that with no conversion in central $Au+Au$
collisions.

Shown in Fig.~(\ref{fig:ratio-p-pie-conversion}) are the $p/\pi^{+}$
and $\bar{p}/\pi^{-}$ ratios in $p+p$ and central $Au+Au$ collisions
with conversion (dotted-dashed line) and without conversion (dotted
line) between quarks and gluons. The ratios of (anti)-proton over
pion with conversion are found to fit data not bad. Net
quark-to-gluon jet conversions effectively increase the total number
of final gluon jets surviving in the fire ball. Therefore the final
fractions of gluon and quark jets contributing to the hadron spectra
will be modified by the net quark-to-gluon conversion. Furthermore,
our numerical results in Fig.~(\ref{fig:fraction}) show that even if
without net quark-to-gluon conversions high $p_T$ proton
(anti-proton) production is dominated by gluon fragmentations
whereas high $p_T$ pion is dominated by quark fragmentations in
central $Au+Au$ collisions. Consequently, larger $p/\pi$ ratios are
found in central $Au+Au$ collisions with conversion than without
conversion, $(\frac{\bar p}{\pi^-})_{AuAu}^{conv.}>(\frac{\bar
p}{\pi^-})_{AuAu}^{no~conv.}$ and
$(\frac{p}{\pi^+})_{AuAu}^{conv.}>(\frac{p}{\pi^+})_{AuAu}^{no~
conv.}$ shown in Fig.~(\ref{fig:ratio-p-pie-conversion}). Similar to
the case in Fig.(\ref{fig:ratio-p-pie}), the ratios of (anti)-proton
over pion in central $Au+Au$ collisions with conversion are smaller
than those in $p+p$ collisions. According to $R_{AA}^{\bar
p}<R_{AA}^{p}\approx R_{AA}^{\pi^+}(=R_{AA}^{\pi^-})$ from
Fig.~(\ref{spec-pie-conversion})(\ref{fig:spec-proton-conversion}),
one can get
\begin{eqnarray}(\frac{p}{\pi^+})^{AuAu}=(\frac{R_{AA}^p}{R_{AA}^{\pi^+}})(\frac{p}{\pi^+})^{pp}
\approx(\frac{p}{\pi^+})^{pp};\\
(\frac{\bar p}{\pi^-})^{AuAu}=(\frac{R_{AA}^{\bar
p}}{R_{AA}^{\pi^-}})(\frac{{\bar p}}{\pi^-})^{pp} <(\frac{{\bar
p}}{\pi^-})^{pp},
\end{eqnarray}
as one can see in Fig.~(\ref{fig:ratio-p-pie-conversion}). $p/\pi^+$
ratios in central $Au+Au$ collisions with conversion are similar to
those in $p+p$ collisions. However $\bar{p}/\pi^-$ ratios in central
$Au+Au$ collisions with conversion are smaller than those in $p+p$
collisions because the number of anti-quark jets is smaller than
that of quark jets and the effect of conversion mechanism is not
obvious for anti-quark jets.

We note that in the conventional jet quenching theory, because a
gluon jet always lose more energy than a quark jet with the same
initial energy in a hot QCD medium (except in the case when both
quark jet and gluon jet lose all of their energies), and at high
$p_T$ $\pi$ meson is dominated by valence quark fragmentation
whereas proton is dominated by gluon fragmentation, the $p/\pi$
ratio at large transverse momentum in $p+p$ collision should be no
less than those in $A+A$ collisions. The two scenarios considered in
this paper will raise the $p/\pi$ ratios in $A+A$ collisions
therefore becoming closer to those in $p+p$, but will never go
beyond. Only in the extreme case in $A+A$ collisions at large $p_T$,
all quark jets are converted into gluon jets, or equivalently there
are only gluon jets, we can get higher $p/\pi$ ratio in $A+A$
collisions at large $p_T$. Though current data at STAR seems to
imply that proton over pion ratio in $Au+Au$ reactions is enhanced
relative to that in $p+p$ collisions, with large error bars at high
$p_T$, we still can not reach a unambiguous conclusion. If in the
future more accurate experimental measurements confirm that at high
$p_T$ region, $p/\pi$ ration in $A+A$ collision is enhanced
significant compared with that in $p+p$ collision, a dramatic change
of our understanding of parton energy loss mechanism will be needed.
Because we focus on $p/\pi$ ratios at the  high $p_T$ region, in our
calculations we have ignored the contribution of parton recombination/coalescence
mechanism in the QGP. We expect this negligence underlines the rather large discrepancy
between our numerical simulations and experimental data in
the intermediate $p_T$ region, where the recombination/coalescence mechanism
in the hot QGP should play a very important role \cite{co1,co2,co3,co4,hwa}.

\section{Summary and discussions}\label{sec:d}

In this paper, we investigate $p/\pi^{+}$ and $\bar{p}/\pi^{-}$
ratios at large transverse momentum in $Au+Au$ collisions with jet
quenching based on an next-to-leading order pQCD parton model. Two
scenarios are considered to explain the anomaly of $p/\pi$ ratios
observed by STAR Collaboration that $p/\pi$ ratios at high $p_T$ in
central $Au+Au$ collisions~\cite{star-auau} approach those in $p+p$
and $d+Au$ collisions~\cite{star-pp}. Firstly, we investigate the
consequence in Scenario I where gluons lose the same amount of
energy in QGP as quarks and demonstrate that it will enhance the
$p/\pi^{+}$ and $\bar{p}/\pi^{-}$ ratios for central $Au+Au$
collisions. Secondly, we explore the effect of strong jet conversion
in Scenario II where a net of quark-to-gluon conversion also result
in similar $p/\pi^{+}$ and $\bar{p}/\pi^{-}$ ratios in central
$Au+Au$ and $p+p$ collisions. In both scenarios, the final number of
gluon jets passing through the hot and dense medium is increased as
compared to the original jet quenching picture, and jet conversion
scenario is favorable when $p/\bar{p}$ ratio is considered because
${\bar p}/p$ in the latter scenario is found to fit data better than
that in the former scenario. This kind of studies should improve our
understanding on the flavor dependence of energy loss and lead us to
a unified picture of jet quenching.

We note that in our approach the effect of energy loss is taken into
account through the medium-modified effective fragmentation
functions and the evolution of the hot nuclear medium is described
by a simple 1+1D Bjorken evolution based on Hard-Sphere geometry.
This approach has given very well descriptions on inclusive pion
suppression~\cite{2-wang}, di-hadron production~\cite{zhanghz} and
photon-tagged hadron production~\cite{gamma}, but several
improvements will be needed to make a more realistic consideration
of parton energy loss in the hot medium created at RHIC. One
direction is to simulate the medium evolution with the 3+1D
hydrodynamics model as in Ref. \cite{fluctuation2,fluctuation3},
which may give corrections to the total parton energy loss though
the effect of local density dropping due to the transverse expansion
of the medium will be offset by the longer path length in the
transverse direction \cite{length} in the QGP phase\footnote{In
other phases of 3+1D hydrodynamical evolution, for instance in the
mixed phase (MP), the path length may shorten with the transverse
expansion of the system\cite{Heinz:2009xj}, and we thank the referee
for pointing out this anomaly of path length variation with the
transverse expansion at the full 3+1D hydrodynamics simulation.}.
Also one could study the in-medium parton shower
\cite{fluctuation2,monte1, monte2} to make a more differential study
of jet quenching than the adoption of the effectively modified
parton fragmentation function. Furthermore a more detailed medium
dependence of jet energy loss can be obtained by introducing the
energy loss probability distributions
\cite{fluctuation2,fluctuation3,fluctuation1,fluctuation4,
fluctuation5}, which will reduce the difference of energy losses of
quark and gluon jets by properly incorporating kinematical
constraints, and thus raise the $p/\pi$ ratios. Last but not the
least, in our calculation we have assumed that the pions and protons
are formed outside of the medium, and the hadronization of pions and
protons in A+A collisions are the same as that in p+p collisions.
Larger $p/\pi$ ratios may be obtained and help to resolve the
$p/\pi$ puzzles observed in A+A collisions at STAR if one assumes
that protons in A+A collisions can be formed inside the medium due
to its larger mass relative to pion , though so far how to model
hadronizations of partons with large $p_T$ in medium and whether it
is distinct from that in vacuum is still an open question and under
hot debate. Nevertheless, intensive theoretical studies will be
needed to see whether we can resolve the puzzle of $p/\pi$ ratios in
a more beautiful and natural way by improving our model and
including all the relevant effects discussed above.

This work was supported by NSFC of China under Projects No. 10825523
and No. 10875052 and No. 10635020, by MOE of China under Projects
No. IRT0624; by MOST of China under Project No. 2008CB317106; and by
MOE and SAFEA of China under Project No. PITDU-B08033.

\end{document}